\documentclass[prl,aps,twocolumn,superscriptaddress,dvipsnames,amssymb,9pt,floatfix]{revtex4}
\usepackage{graphicx}
\usepackage{amsmath,amssymb,amsthm,amsfonts,mathptmx}
\usepackage{color}
\usepackage[matrix,frame,arrow]{xypic}
\usepackage{times}
\usepackage{algorithm}
\floatname{algorithm}{Protocol}
\theoremstyle{plain}

\theoremstyle{definition}

\newcommand{\ket}[1]{|#1\rangle}

\DeclareMathOperator{\CZ}{CZ}
\DeclareMathOperator{\Hd}{H}

\begin{document}
\title{Overcoming efficiency constraints on blind quantum computation}

\author{Carlos A. P\'erez-Delgado}
\affiliation{Singapore University of Technology and Design, 20 Dover Drive, Singapore 138682}
\author{Joseph F. Fitzsimons}\email{joseph_fitzsimons@sutd.edu.sg}
\affiliation{Singapore University of Technology and Design, 20 Dover Drive, Singapore 138682}
\affiliation{Centre for Quantum Technologies, National University of Singapore, 3 Science Drive 2, Singapore 117543}
\begin{abstract}
Blind quantum computation allows a user to delegate a computation to an untrusted server while keeping the computation hidden. A number of recent works have sought to establish bounds on the communication requirements necessary to implement blind computation, and a bound based on the no-programming theorem of Nielsen and Chuang has emerged as a natural limiting factor. Here we show that this constraints only hold in limited scenarios and show how to overcome it using a method based on iterated gate-teleportations. We present our results as a family of protocols, with varying degrees of computational-ability requirements on the client. Certain protocols in this family exponentially outperform previously known schemes in terms of total communication. The approach presented here can be adapted to other distributed computing protocols to reduce communication requirements.
\end{abstract}
\maketitle

Blind quantum computation is a cryptographic task whereby a client seeks to hide a delegated computation from the server implementing the computation. A number of protocols for blind computation have been discovered exhibiting either information theoretic security \cite{bfk09,Dorit,BKBFZW11,FM12,fk12,Morimae12,morimae2013blind,barz2013experimental,mpf13,SKM13}, for which it can be proven that a cheating server cannot learn anything about the computation being performed based on purely information theoretic grounds, or cheat-sensitivity  \cite{gmmr13}, whereby a  cheating server can be detected even though the computation cannot be  kept hidden from it. A range of capabilities for the client have also been considered, from the ability to prepare or measure individual single qubit states \cite{bfk09, fk12, Morimae12}, to the ability to perform universal computation on fixed size systems \cite{Dorit}. Recently work has sought to unify this disparate family of protocols in terms of security definitions \cite{dunjko2013composable} and in terms of resource accounting \cite{mpf13,gmmr13}.

Recently, Giovannetti, Maccone, Morimae and Rudolph proposed a novel cheat-sensitive protocol for blind quantum computation, for which the total communication required scales optimally for their specific setting \cite{gmmr13}. However, their optimality argument applies only to the case where the client is restricted to preparing and performing projective measurements on single qubits in two bases. In this case, they argued---based on the \emph{no-programming} theorem of Nielsen and Chuang \cite{nc97}---that $\Omega(J\log_2 G)$ qubits must be exchanged between client and server, where $J$ is the total number of gates performed, and $G$ is the cardinality of the gate set. A slightly weaker bound than that provided by the no-programming theorem can be shown for any protocol where the client is restricted to preparing or measuring qudits of dimension $D$ in one of $B$ bases, by counting the number of possible branches of the protocol: the client's actions amount to a series of $s$ preparations and measurements, since any classical communication can also be viewed in this way. As the client has no memory, preparations correspond to the transmission of a qudit to the server, and measurements correspond to the transmission of a qudit from the server. As such, $s$ qudits in total are exchanged. In order for the client's actions to determine which of $N_C$ possible computations is performed, it is necessary to have $s \geq \frac{\log N_C}{\log Bd}$ by the pigeon-hole principle. Hence, in this setting, blind computation of a circuit of depth $J$ consisting of gates independently chosen from a set of cardinality $G$ requires an exchange of at least $\frac{J\log G}{\log Bd}$ qudits. While for any fixed choice of $d$ and $B$ this approach yields a bound similar to the no-programming theorem, it is more widely applicable. As such, it is tempting to conjecture that the no-programming bound of $\Omega(J\log_2 G)$ on the number of bits/qubits exchanged applies to any approach to blind quantum computation. In this paper, however, we show that such an efficiency constraint can be overcome if the client is allowed to prepare arbitrary single or multi-qubit states. The protocols we construct are not only more efficient than previous blind computation protocols, but require less communication than is required to classically describe the delegated computation.

\begin{figure}[t!]
\includegraphics[width=0.8\columnwidth]{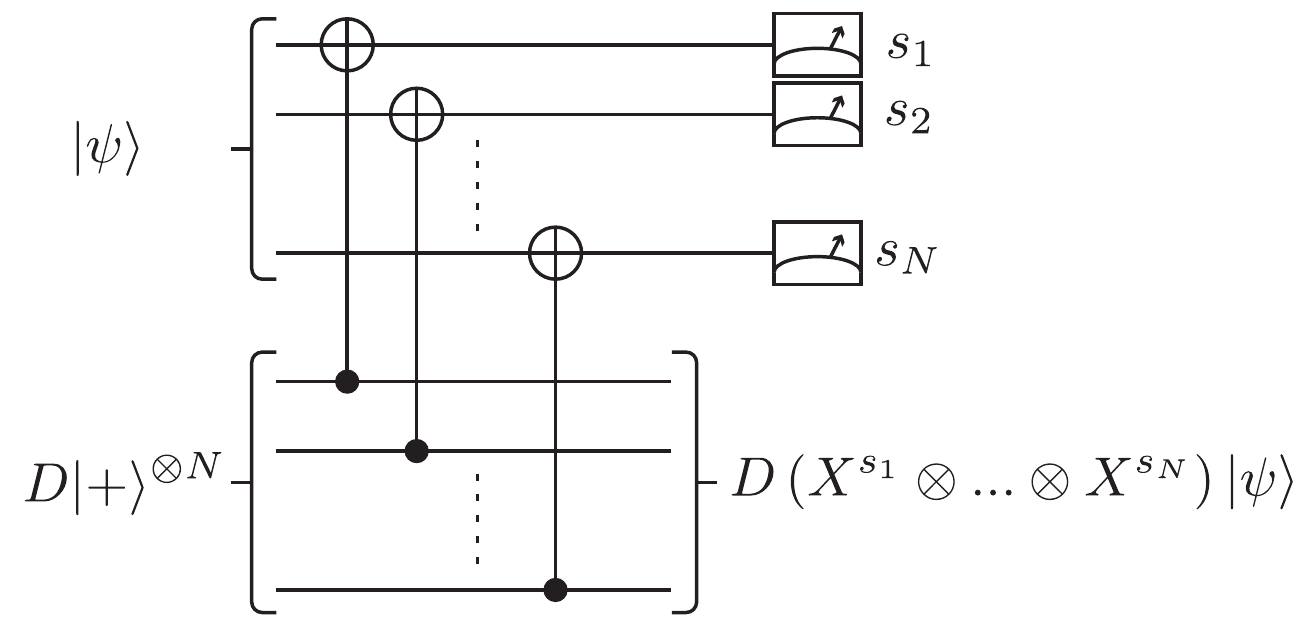}
\caption{Gate teleportation procedure. The top set of wires corresponds to register $R$, while the bottom set correspond to $R'$.\label{fig:tele}}
\end{figure}

Before presenting the main protocol, we first present a precursor protocol which forms the main building block of our final protocol. Our approach is based on gate teleportation \cite{gottesman}, but differs from standard usages in that instead of directly correcting errors induced by teleportation byproducts, we make use of additional gate teleportation steps to correct the state of the system. This avoids the need to provide a classical description of the correction operator, leading to a saving in the total communication cost of the protocol, in turn allowing our protocol to avoid the lower bound on communication of $J\log_2 G$ which results from a naive application of the no-programming theorem. 

Consider the set $\mathcal{D}_{m,l}$ of all diagonal operators acting on $m$ qubits of the form $\exp( i \sum_{j\in \{0,1\}^m} \theta_j Z^{j_1} \otimes Z^{j_2} \otimes \ldots \otimes Z^{j_m})$, with $\theta_j \in \{\frac{r \pi}{2^l}| r\in \{0,1,2,3,\ldots,2^l-1\} \}$. Our approach allows Alice to successfully teleport any given operator $D \in \mathcal{D}_{m,l}$ to Bob in at most $l$ steps, each involving the transmission of $m$ qubits. This gives a total cost of $O(m l)$. Compare this to any setting where the no-programming theorem applies, which sets a minimum of $\Omega(l 2^m)$ qubits to be transmitted.

\begin{algorithm}[t]
\caption{Iterated Teleportation}\label{prot:1}
\begin{description}
\item[Alice's input:] Gate $D \in \mathcal{D}_{m,l}$ to be teleported.
\item[Bob's input:] Initial state $\ket{\psi}$, in register $R$.
\item[Output:] The state $\overline{X} D \ket{\psi} $ in Bob's register $R$, where $\overline{X}$ is a tensor product of the Pauli-$X$ operator and the identity, known both to Bob and Alice.
\item[Steps:]
\end{description}
\begin{enumerate}
\item Set $D_1 = D$.
\item For $1 \leq \ell \leq l$
\begin{enumerate}
\item\label{goto:1} Alice prepares the state $D_{\ell} \ket{+}^{\otimes m}$, and sends it to Bob, who stores it in register $R'$.
\item Bob applies the teleportation procedure depicted in Fig.\ \ref{fig:tele} obtaining measurement results $s_1^{(\ell)}, \ldots, s_m^{(\ell)}$. He sends the measurement results to Alice.
\item Bob swaps the contents of register $R$ and $R'$.
\item\label{end:1}  Set $D_{\ell+1} = \overline{X}_{\ell}D_{\ell}\overline{X}_{\ell}D_{\ell}^\dagger,$ where $\overline{X}_{\ell} = \bigotimes_i^m X^{s_i^{(\ell)}}$.
\end{enumerate}
\item Set $\overline{X} = \left(\prod_{\ell=1}^l \overline{X}_{\ell} \right)$. Bob now has the desired state $\overline{X}D \ket{\psi} $ in register $R$.
\end{enumerate}
\end{algorithm}

We will assume that Bob's system contains two registers $R$ and $R'$.
The multi-qubit gate teleportation circuit we use is depicted in Fig.\ \ref{fig:tele}. This procedure is formalised in Prot.\ \ref{prot:1}. 
Note that at the end of Prot.\ \ref{prot:1}, Bob is in possession of the desired output state, up to a series of Pauli-$X$ corrections, which he can perform himself---in this non-blind version. 
Before discussing a blind version of Prot.\ \ref{prot:1} we show that this protocol, if followed by both Alice and Bob, does indeed yield the correct output \footnote{A minor modification of Prot.\ \ref{prot:1} involves Alice and Bob halting the protocol as soon as Bob measures all zeroes. In this case, no further corrections are necessary, and the protocol can conclude with the correct output state. In the case where $2^m < l$ this leads to an average-case communication cost which is independent of $l$.}.

 We will begin by examining the effect of an iteration of the main loop (Steps \ref{goto:1} through \ref{end:1}) on an arbitrary input state $\ket{\psi}$ in register $R$. Each iteration serves to implement a gate teleportation so that an input state $\ket{\psi_{\ell}}$ is transformed to $\ket{\psi_{\ell+1}} =  D_{\ell} \overline{X}_{\ell} \ket{\psi_{\ell}}$. Thus, 
\begin{equation}
\ket{\psi_{l}} =  
\left(\prod_{\ell=1}^l 
 D_{\ell} \overline{X}_{\ell} \right) 
 \ket{\psi},
\end{equation}
where the product operator is used to denote that the left to right ordering is from highest to lowest value of $\ell$. Note that if an operator $D \in \mathcal{D}_{m,t}$ then $\left(\bigotimes_{k=1}^n X^{a_k}\right) D \left(\bigotimes_{k=1}^n X^{a_k}\right) D^\dagger \in \mathcal{D}_{m, t-1}$ for any choice of variables $a_k \in \{0,1\}$. Thus, for any $\ell$, we have $D_{\ell} \in \mathcal{D}_{m,l-\ell+1}$. Since $\mathcal{D}_{m,1}$ corresponds to the set of tensor products of $Z$ and the identity, $D_{l} \overline{X}_{l} = \pm \overline{X}_{l} D_{l}$. Thus, up to a global phase, we have
\begin{equation}
\ket{\psi_{l}} = 
\overline{X}_{l}
D_{l} \left(\prod_{\ell=1}^{l-1} D_{\ell} \overline{X}_{\ell} \right)
\ket{\psi_{j}},
\end{equation}
which collapses telescopically, substituting in the definition of $\overline{D}_{\ell}$, to yield
\begin{equation}
\ket{\psi_{l}} = \left(\prod_{\ell=1}^l \overline{X}_{\ell} \right) D_{1} \ket{\psi_{j}}.
\end{equation}
Setting $\overline{X} = \left(\prod_{\ell=1}^l \overline{X}_{\ell} \right)$ completes the proof. Note that, at this stage, Bob can correct his state by applying Pauli-$X$ to his qubits as appropriate, without knowing the teleported gate, and without any further assistance or communication from Alice.

Next we present a blind version of Prot.\ \ref{prot:1}. This procedure allows for the same functionality as the previous protocol, enabling Alice and Bob to perform gate teleportation of a gate encoded by Alice, while additionally ensuring that the gate remains unknown to Bob. The procedure for accomplishing this is presented in Prot.\ \ref{prot:2}.

\begin{algorithm}[t]
\caption{Blind Iterated Teleportation}\label{prot:2}
\begin{description}
\item[Alice's input:] Gate $D \in \mathcal{D}_{m,l}$ to be teleported.
\item[Bob's input:] Initial state $\ket{\psi}$, in register $R$.
\item[Output:] The state $\overline{Z}\overline{X} D \ket{\psi} $ in Bob's register $R$, where $\overline{Z}$ ($\overline{X}$) is a tensor product of the Pauli-$Z$ (Pauli-$X$) operator and the identity, and $\overline{Z}$ is Alice's encryption key,  known only to her.
\item[Steps:]
\end{description}
\begin{enumerate}
\item Set $D_1 = D$.
\item For $1 \leq \ell \leq l$
\begin{enumerate}
\item\label{goto:2} Alice prepares the state $\overline{Z}_{\ell}\overline{Z}_{\ell-1} D_\ell \ket{+}^{\otimes m}$, where $\overline{Z}_{\ell} = \bigotimes_{k=1}^m Z^{r_k^{(\ell)}}$, where $r_k^{(\ell)}$, $1 \leq \ell \leq x$ are uniformly random bits, and $\overline{Z}_{0} = I$. She transmits it to Bob, who stores it in register $R'$.
\item Bob applies the teleportation procedure depicted in Fig.\ \ref{fig:tele} obtaining measurement results $s_1^{(\ell)}, \ldots, s_m^{(\ell)}$. He sends the measurement results to Alice.
\item Bob swaps the contents of register $R$ and $R'$.

\item\label{end:2}  Set $D_{\ell+1} = \overline{X}_{\ell}D_{\ell}\overline{X}_{\ell}D_{\ell}^\dagger,$ where $\overline{X}_{\ell} = \bigotimes_i^m X^{s_i^{(\ell)}}$.
\end{enumerate}
\item Set $\overline{X} = \left(\prod_{\ell=1}^l \overline{X}_{\ell} \right)$, and  $\overline{Z} = \overline{Z}_l$. Bob now has the desired state  $\overline{Z}\overline{X} D \ket{\psi} $ in register $R$.\end{enumerate}
\end{algorithm}

The proof of correctness of Prot.\ \ref{prot:2} is similar to that of Prot.\  \ref{prot:1}. The only difference is that now each iteration adds a series of random Pauli-$Z$ operators. With each iteration an input state $\ket{\psi_{\ell}}$ is transformed to $\ket{\psi_{\ell+1}} =  \overline{Z}_{\ell}\overline{Z}_{\ell-1} D_{\ell} \overline{X}_{\ell} \ket{\psi_{\ell}}$. Thus, 
\begin{equation}
\ket{\psi_{l}} =  
\left(\prod_{\ell=1}^l 
\overline{Z}_{\ell}\overline{Z}_{\ell-1}  D_{\ell} \overline{X}_{\ell} \right) 
 \ket{\psi},
\end{equation}
where, again, the product operator is used to denote that the left to right ordering is from highest to lowest value of $\ell$. As before, we use the property of $\mathcal{D}_{m,1}$, and the definition of $D_{\ell} $ to telescopically collapse the last equation to
\begin{equation}
\ket{\psi_{l}} = \overline{Z}_{l} \left(\prod_{\ell=1}^l \overline{X}_{\ell} \right) D_{1} \ket{\psi_{j}}.
\end{equation}

Now we turn to the blindness of the protocol.
The only information transmitted from Alice to Bob are  the set of quantum states $\ket{\phi_{\ell}} = \overline{Z}_{\ell}\overline{Z}_{\ell-1} D_\ell \ket{0}^{\otimes m}$. Note that only $\ket{\phi_{l}}_k$ is a function of $r_{k}^{(l)}$ for $1\leq k\leq m$, and that as these values are unknown to Bob, this state is necessarily the maximally mixed state of $m$ qubits, and hence independent of the values of all other $r_{k}^{(\ell)}$ when  $\ell \neq l$. Hence $\ket{\phi_{l-1}}$ is the only quantum state dependent on $r_{k}^{(l-1)}$ for $1\leq k\leq m$, and must similarly be in a maximally mixed state. We can apply this argument recursively, implying that every state sent from Alice to Bob is in the maximally mixed state due to the unknown values of $\{r_{k}^{\ell}| 1\leq k \leq m, 1\leq\ell\leq l\}$, which serve as Alice's key. As the joint state of all these messages are fixed to the maximally mixed state and hence independent of the computation, the only parameters leaked to Bob are $l$ and $m$.

We are now in a position to present a complete protocol for universal blind quantum computation. Rather than present a single protocol,
we introduce a family of protocols which are parameterised by an integer $m$, corresponding to the size of the client's system. During intermediate steps  Alice will instruct Bob  to perform $m$-qubit gates via gate-teleportation, using Prot.\ \ref{prot:2}. The purpose of studying protocols with varying parameter $m$ is the following. When $m =1$, we obtain a protocol with the most modest computational requirements on  Alice---she need only be able to prepare and send single qubits in a finite set of states---while still outperforming previous protocols. On the other hand, when $m$ is proportional to the number of qubits used in the computation, $n$, the protocol achieves an exponential (in $n$) separation in total communication from the naive limit implied by the no-programming theorem.

 \begin{figure}
 \includegraphics[width=\columnwidth]{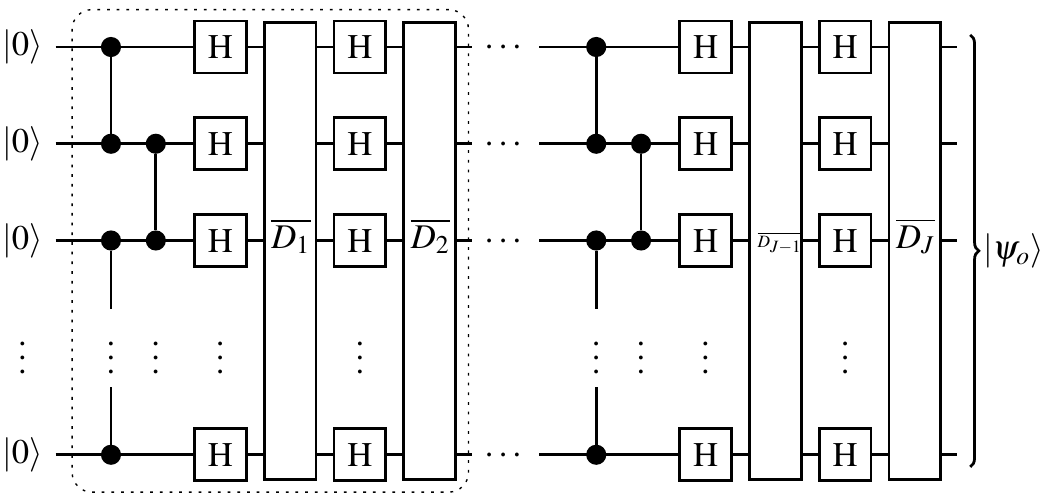}
 \caption{Blind Quantum Computation with Teleportation Protocol. The dotted-line square shows the repeating pattern of operations.\label{fig:prot}}
 \end{figure}

The protocol proceeds in phases. During the $j$th \emph{teleportation} phase, Alice will use gate teleportation to send the desired gates to Bob. Without loss of generality we assume that $m$ divides $n$ \footnote{This may always be done, since Alice can pad her input with unused ancilla qubits.}, such that $n = P m$, for some integer $P$. Then for each contiguous set of $m$ qubits Alice will teleport an operator $D_{j,p}$, to Bob, for $1 \leq p \leq P$, who will then apply it to the qubits labelled $(p-1) m + 1$ through $pm$ of his current state in memory. Thus, if at the beginning of the $j$th phase Bob's register is in the state $\ket{\psi_j}$, by the end of the phase it will be in state $\overline{D_j}\ket{\psi_j}$, where $\overline{D_j} = \otimes_{p =1}^P D_{j,p}$.

As the set of gates which may be implemented by gate teleportation do not form a universal gate set, the scheme we present here leverages fixed gates implemented by Bob to bring about universality in a completely blind manner, as follows. Interspersed with the operations that Alice teleports, Bob will also apply the operation $\overline{\CZ} = \prod_{i=1}^{n-1}\CZ(i,i+1)$, where $\CZ(i,i+1)$ is the controlled-$Z$ operator acting on qubits $i$ and $i+1$;  as well as the operator $\overline{\Hd} = \Hd^{\otimes n}$, where $\Hd$ is the usual Hadamard operator. The order of phases of the protocol is as follows. The first step consists of controlled-$Z$ operators, followed by a Hadamard step, then a teleportation phase, then another Hadamard phase, followed by a second teleportation phase. Then, the pattern repeats itself until $J$ teleportation phases have been achieved. To simplify the analysis, and without loss of generality, we assume $J$ is even. This set of operations forms a universal set of gates for quantum computation for any $m$, and any $x \geq 2$ (see for example \cite{joe2007}). See Fig. \ref{fig:prot} for a schematic diagram and Prot.\ \ref{prot:3} for formal presentation of the protocol.

\begin{algorithm}[t]
\caption{General Iterated Teleportation Blind Quantum Computation}\label{prot:3}

\begin{enumerate}
\item Alice chooses a depth $J$ and a set of diagonal operations $\overline{D}_j = \bigotimes_{p =1}^P D_{j,p}, \, D_{j,p} \in \mathcal{D}_{m,x}$,
 where $P = n / m$ and $n$ is the number of qubits used in the computation, such that her target computation is given by the measurement of 
$\overline{\Hd} \overline{D}_J \overline{\Hd} \overline{D}_{J-1}  \overline{\Hd}  \overline{\CZ} \dots  \overline{\CZ} \overline{D_2} \overline{\Hd} \overline{D}_1 \ket{+}^{\otimes n}$ in the computational basis.

\item Alice produces the state $\overline{Z}_1\overline{D}_1 \ket{+}^{\otimes n}$, 
where $\overline{Z}_{1} = \bigotimes_{k=1}^m Z^{r_k^{(1)}}$, where each $r_k^{(1)}$ is chosen uniformly at random from the set $\{0,1\}$, and transmits $J$ and this state to Bob, who stores the quantum state in register $R$.
\item For $2\leq j \leq J$
\begin{enumerate}
\item If $j \equiv 1 \mkern-9mu \mod 2$, then Bob applies $\overline{\CZ}$ to register $R$.
\item Bob applies $\overline{\Hd}$ to register $R$.
\item\label{step:f} For $1 \leq p \leq P$ 
\begin{enumerate}
\item Alice calculates the operator $f_{j,p}(D_{j,p})$, where the function $f_{j,p}$ is defined in the main text in Eq. \ref{eq:general_corr}.

\item Alice and Bob engage in Prot. \ref{prot:2} using $f_{j,p}(D_{j,p})$ as Alice's target gate, and Bob's qubits $(p-1)m$ through $pm$ as the target register.

\item Alice keeps a record of the operator $\overline{X}_{j,p}$, the teleportation byproduct resulting from Prot. \ref{prot:2}, and $\overline{Z}_{j,p}$ her encryption key.
\end{enumerate}
\item Alice calculates the operators
\begin{equation}
\overline{X}_{j} = \bigotimes_p^{n/m} \overline{X}_{j,p}, \quad
\overline{Z}_{j} = \bigotimes_p^{n/m} \overline{Z}_{j,p},
\end{equation}
and keeps a record of them.
\end{enumerate}

\item Finally, Bob measures his resulting state in the $X$ basis, and sends the measurement outcomes $m_1 \ldots m_n$ to Alice. Alice computes each output bit for the computation as $o_k = m_k \oplus r_k^{{J}}$, where $\overline{Z}_{J} = \bigotimes_k Z^{r_k^{{J}}}$.
\end{enumerate}
\end{algorithm}

The correct operation of the protocol depends on the proper definition of the function $f_j$ used in Step \ref{step:f}. This function is meant to correct and remove the $X$ errors and the $Z$ obfuscation operators introduced in previous steps. Before giving a general definition of $f_j$, lets first consider a simplified version of the protocol where $m = n$ and the phases of controlled-$Z$ operators have been subsumed into the diagonal operator teleportation phases. Hence, the protocol simplifies into a series teleported gate phases followed by a layer of Hadamard gates. The output of the protocol is then given by
\begin{equation}\label{eq:simplified}
\ket{\psi_o} =  \prod_{j=1}^J \left( \overline{Z}_j\overline{X}_j f_j(D_j) \overline{\Hd} \right)  \ket{0}^{\otimes n},
\end{equation}
where the product operator is used to denote that the left to right ordering is from highest to lowest value of $j$.

Because there is a layer of Hadamard gates in between every teleportation stage, the $Z$ operator byproducts are turned into $X$ operators, and \emph{vice versa}, before the the next teleportation. Since Alice can only implement diagonal gates using Prot. \ref{prot:2}, she can only correct the $X$ byproducts of the previous stage. She can, however, conjugate her current gate with the previous $Z$ operators, so as to commute that operator forward, so that it can be corrected in the following teleportation stage. In this case, $f_j$ is given by:
\begin{equation}
f_j(D) = \overline{\Hd}\overline{Z}_{j-1}\overline{\Hd} D \overline{Z}_{j-2}
\overline{\Hd} \overline{X}_{j-1} \overline{Z}_{j-1} \overline{\Hd},
\end{equation}
where $\overline{Z}_{j} = \overline{X}_{j} = I$ for all $j < 1$.
It is straightforward to verify from the definition above that $f_j$ maps $\mathcal{D}_{m,l}$ onto itself. Now, substituting into Eq. \ref{eq:simplified} we get
\begin{equation}
\ket{\psi_o} = \overline{Z}_J\overline{X}_J \prod_{j=1}^J \left( D_j \overline{\Hd} \right)  \ket{0}^{\otimes n}.
\end{equation}

From this state, Alice can get the correct output for computation by having Bob measure in the $X$ basis, and sending her the output. After this, she uses her decryption key. 

The analysis of the full protocol is slightly more involved due to the (re-)introduction of the $\CZ$ gates, since these affect the propagation of $X$ byproducts. Fortunately, errors propagate only \emph{once}, to the nearest neighbours before they can be corrected. To see this, note that in between every two $\CZ$ gate stages, there are two sequences of Hadamard operators followed by  gate teleportations. Hence, any error that cannot be fixed in the gate-teleportation phase immediately preceding the $\CZ$ stage, can be corrected in the subsequent phase. In order to define the general correction function $f_{j,p}$, let $x_{j,k}$ and $y_{j,k}$ be such that
\begin{equation}\label{eq:fj0}
\overline{X}_{j} = \bigotimes_{k=1}^n X^{x_{j,k}},\quad \mbox{and} \quad 
\overline{Z}_{j} = \bigotimes_{k=1}^n Z^{z_{j,k}}.
\end{equation}

For even $j$ we take $\chi_{j,k} = z_{j-1,k}$ and $\zeta_{j,k} = z_{j-2,k} + 
x_{j-1,k} + \sum_{t \in \{-1,1\}}\left(z_{j-3,k+t} + x_{j-2,k+t}\right)$, 
and for odd $j$ we take $\chi_{j,k} = z_{j-1,k} + \sum_{t \in \{-1,1\}}\left(z_{j-2,k+t} + x_{j-1,k+t}\right)$ and $\zeta_{j,k} = z_{j-2,k} + x_{j-1,k}$.
Then, finally, for all $j$ and $p$ we define 
\begin{equation}\label{eq:general_corr}
f_{j,p}(D) = \left( \bigotimes_{k} X^{\chi_{j,k}}\right) D \left( \bigotimes_{k} Z^{\zeta_{j,k}} X^{\chi_{j,k}}\right),
\end{equation}
where the tensor products are taken over values of $k$ ranging from $(p-1)m + 1$ to $pm$.

The correctness of the general protocol follows from a similar argument to that of the special case previously considered. The output of the general protocol is given by
\begin{equation}\label{eq:general}
\ket{\psi_o} =  \prod_{j=1}^J \left( \overline{Z}_j\overline{X}_j \left( \bigotimes_{p=1}^{n/m} f_{j,p}(D_{j,p}) \right) \overline{\Hd} \overline{\CZ}^j \right)  \ket{0}^{\otimes n},
\end{equation}
where the product operator is used to denote that the left to right ordering is from highest to lowest value of $j$. Substituting Eq. \ref{eq:general_corr} into Eq. \ref{eq:general}, with some elementary algebra, gives
\begin{equation}
\ket{\psi_o} = \overline{Z}_J\overline{X}_J \prod_{j=1}^J \left( D_j \overline{\Hd}  \overline{\CZ}^j\right)  \ket{0}^{\otimes n},
\end{equation}
as required.

The proof of blindness of the main protocol follows directly from the proof of blindness of Prot. \ref{prot:2}.

Setting the parameter $m$ equal to $n$, the above protocol requires that Alice transmit exactly $Jx$ different $n$ qubit states to Bob, and hence requires only a total of $nJx$ qubits to be sent from Alice to Bob, and $nJx$ classical bits to be sent from Bob to Alice. However, this protocol implements $J$ unknown operations, each of which can be drawn arbitrarily from the set $\mathcal{D}_k$ which has cardinality $(2^k)^{(2^n - 1)}$. Any scheme to which the no-programming theorem applies, such as that in \cite{gmmr13}, would require that at least $Jk(2^n - 1)$ qubits or bits be communicated, and hence the protocol presented here is exponentially more efficient than than previous schemes.

Setting $m$ equal to $n$ requires Alice to prepare large entangled states, which may be undesirable in realistic settings. Setting $m$ to a small constant, gives a protocol that is at least as easily implementable as previous ones in terms of resources needed on Alice's side, while still being universal for quantum computation, and offering a communication advantage.

The approach of iterated teleportation introduced in this comment can be applied to other measurement-based blind computation protocols (such as \cite{bfk09}, \cite{fk12} and \cite{mpf13}) to achieve smaller advantages over the no-programming theorem bound. It can be used outside of the blind computation setting, to reduce communication requirements in other delegated computation scenarios via Prot. \ref{prot:1}.The notion of blindness used here is compatible with requirements for composable security in the abstract cryptography framework \cite{dunjko2013composable}, meaning that there is no compromise in security. 

This material is based on research funded by the Singapore National Research Foundation under NRF Award NRF-NRFF2013-01. We thank Lorenzo Maccone, Tomoyuki Morimae and Terry Rudolph for helpful comments on the manuscript.

\bibliographystyle{apsrev}
\bibliography{Comment}

\end{document}